\documentclass[%
 reprint,nobibnotes,
 amsmath,amssymb,
 aps,prd,
12pts]{revtex4-1}

\usepackage{dcolumn} 
\usepackage{bm} 
\usepackage{graphicx,tensor}
\usepackage{float}
\usepackage[backref=none]{hyperref} 
\usepackage[usenames,dvipsnames]{color}
\usepackage{fancyhdr}

\definecolor{MyDarkBlue}{rgb}{0,0.1,0.7}

\hypersetup{pdfborder={0 0 0},colorlinks,breaklinks=true,
  urlcolor={MyDarkBlue},citecolor={MyDarkBlue},linkcolor={MyDarkBlue}}

\begin{document}

\title{What does a measurement of mass and/or radius of a neutron star constrain: Equation of state or gravity?}
\author{Kaz{\i}m Yavuz Ek\c{s}i, Can G\"{u}ng\"{o}r, Murat Metehan T\"{u}rko\u{g}lu}

\affiliation{Istanbul Technical University, Faculty of Science and Letters,  \\
Department of Physics, 34469, Maslak, Istanbul, Turkey}

\date[]{Received 29 November 2013; published 6 March 2014}

\email{eksi@itu.edu.tr}

\begin{abstract}
Neutron stars are thought to be excellent laboratories for determining
the equation of state (EoS) of cold dense matter. Their strong gravity
suggests that they can also be used to constrain gravity models.
The two observables of neutron stars---mass and radius (M-R)---both depend on the choice of EoS and
relativistic gravity, meaning that neutron stars cannot be simultaneously
good laboratories for both of these questions. A measurement of mass and/or radius
would constrain the less well known physics input.
The most common assumption---namely, that M-R measurements can be used to constrain 
EoS---presumes general relativity (GR) is the ultimate model of gravity in the classical regime.
We calculate the radial profile of compactness and curvature
(square root of the full contraction of the Weyl tensor) within a neutron star
and determine the domain not probed by the Solar System tests of GR.
We find that, except for a tiny sphere of radius less than a millimeter at the center,
the curvature is several orders of magnitude
above the values present in Solar System tests. The compactness is beyond the solar
surface value for $r>10$ m, and increases by 5 orders of magnitude towards the surface.
With the density being only an order of magnitude
higher than that probed by nuclear scattering experiments,
our results suggest that the
employment of GR as the theory of gravity describing the hydrostatic equilibrium
of the neutron stars is a rather remarkable extrapolation from the regime of tested validity, as opposed to that of EoS models.
 Our larger ignorance of gravity
within neutron stars suggests that a measurement of mass and/or radius constrains gravity
rather than the EoS, and given that the EoS has yet to be determined by nucleon scattering experiments, M-R measurements cannot tightly constrain the gravity models either. Near the surface the curvature and compactness attain their largest
values,  while the EoS in this region is fairly well known. This renders the crust as the best site to look for deviations from GR. \\
\vspace{1cm}
\doi{10.1103/PhysRevD.89.063003} \hspace{2.7cm} PACS numbers: 97.60.Jd, 04.40.Dg, 04.80.Cc

\end{abstract}


\maketitle


\section{Introduction}
\label{intro}

\thispagestyle{fancyplain}

Neutron stars with their diversity of astrophysical manifestations provide many challenging problems
that make them interesting in their own right.
Two aspects of neutron stars, however, make these objects especially important for fundamental physics:
(i) the high densities prevailing at the core of neutron stars make them possible laboratories
for determining the equation of state (EoS) of the ultimate ground state of cold catalyzed dense matter
(e.g., Refs. \citep []{guv+1820,guv+1608,guv+1745,oze+10,guv13,ste+10,oze+12,neu+12,sul+11}; see Refs. \citep[]{lat07,lat12,han+07} for reviews);
(ii) the strong gravitational fields of neutron stars allow them to be used as tools for testing or constraining alternatives \citep[]{har98,sot04,yag+13,hor11,sot12,las+08,pan+11} or modifications \citep[]{coo+10,ara+11,del+12,odd+23,odd+24} of general relativity (GR)
(see Refs. \citep[]{wil06,wil09,psa08} for reviews). Neutron stars cannot simultaneously
be good test beds for both questions (see, e.g., Refs. \citep[]{wen+11}).
The often-mentioned notion that neutron stars are excellent labs to constrain
the EoS presumes that GR is the ultimate theory of gravity at the classical level.
This seems to be a safe assumption given that the parametrized post-Newtonian (PPN)
parameter $\gamma$ (a measure of curvature) has been tested
with a precision of $10^{-5}$ \citep{ber+03,sha+04} and found to be consistent with the prediction
of GR, $\gamma=1$.
As argued in the excellent review of Psaltis \citep{psa08}, however, these Solar System tests constrain
only weak-field gravity; gravity in the strong-field regime is largely unconstrained by observations.
The differences between GR and its alternatives/modifications can become prominent in the strong
gravitational fields of neutron stars \citep{ded03}.

The mean density of a typical neutron star is $\bar{\rho}=3M_{\ast}/4\pi R_{\ast}^3 = 0.67\times 10^{15}M_{1.4}R_{6}^{-3}$ g cm$^{-3}$ ($M_{1.4}=M_{\ast}/1.4M_{\odot}$ and $R_{6}=R_{\ast}/10^6\, {\rm cm}$).
This is  an order of magnitude higher than the densities probed by nucleon scattering experiments
\citep[e.g.][]{dan+02}.  At the surface of the star the compactness
$\eta_{\ast} \equiv 2GM_{\ast}/R_{\ast}c^2 = 0.416\,M_{1.4}R_{6}^{-1}$
and ``curvature''
${\cal K}_{\ast} \equiv 4\sqrt{3}GM_{\ast}/R_{\ast}^3 c^2 = 1.18\times 10^{-12}\,M_{1.4}R_{6}^{-3}$~cm$^{-2}$ (see Sec. \ref{sec-method} for the definition of curvature), the two measures of the gravitational field strength \citep{psa08}, are 5 and 14 orders of magnitudes larger, respectively, than their counterparts in the Solar System ($\eta_{\odot} \equiv 2GM_{\odot}/R_{\odot}c^2 = 4.27\times 10^{-6}$
and ${\cal K}_{\odot}  \equiv 4\sqrt{3}GM_{\odot}/R_{\odot}^3 c^2 =3.06\times 10^{-27}$~cm$^{-2}$).
This large unconstrained regime in gravity compared to that in the EoS may not immediately imply that a measurement of the mass and/or radius (M-R) of a neutron star  would constrain gravity models rather than the EoS of dense matter:
the density of a neutron star is nearly constant at the core and drops by 15 orders of magnitude near the surface meaning that a large fraction of the mass and radius of the neutron star might be determined by the unconstrained
EoS regime, while the radially increasing compactness and curvature within the star could be in a well-tested regime for most parts of the star. Such an argument is probably the reason for the persistence of the general consent within the astrophysics community that a sufficiently accurate M-R measurement would constrain the EoS at the core of neutron stars rather than gravity. Yet another reason could be that GR is singled out among its alternatives as being the simplest valid model of gravity, whereas there is no EoS that is equally as prominent. To decide which---gravity or the EoS---is better understood, it is necessary to  quantify where within the star the strength of gravity is beyond the values probed in Solar System tests.

The two future X-ray missions---NASA's Neutron star Interior Composition ExploreR (NICER) \citep{gen+12} and ESA's the Large Observatory For X-ray Timing (LOFT) \citep{fer+12}---are expected to measure the mass and radius of neutron stars \citep{lo+13,psa+13}. The measurement of the mass and/or radius of a neutron star would constrain the less well known physics input more than the other. Whether results from these missions would constrain the EoS or gravity models depends on which of these physics inputs is less well known within neutron stars. The purpose of this work is to quantify the range of unconstrained gravity within neutron stars in the framework of GR. This clarifies how much the M-R relations obtained within the framework of GR extrapolate its usage from its well-tested regime.
In Sec. \ref{sec-method} we present the hydrostatic equilibrium equations and the definitions of the compactness and curvature that we employ. In Sec. \ref{sec-results} we present our results showing the radial profiles of the compactness and curvature within the neutron star. Finally, in Sec. \ref{sec-discuss} we discuss our results. We present the derivations of the curvature scalars in an Appendix.

\section{METHOD}
\label{sec-method}

The model for gravity determines the macroscopic hydrostatic equilibrium equations whereas the EoS is the microscopic physics input. The mass and the radius of the star (the two obvious observables) are determined by the choice of both the gravity model and the EoS.
In GR the hydrostatic equilibrium of a star is described by the Tolman-Oppenheimer-Volkov (TOV) equations,
\begin{equation}
\frac{dP}{dr} = -\frac{Gm\rho}{r^2}
                 \left( 1+ \frac{P}{\rho c^2}\right)
                 \left(1+\frac{4\pi r^3 P}{mc^2} \right)
                 \left(1-\frac{2Gm}{rc^2}\right)^{-1}
\label{tov1}
\end{equation}
and
\begin{equation}
\frac{dm}{dr}=4\pi r^2 \rho
\label{tov2}
\end{equation}
\citep{tol39,opp39} where $\rho=\rho(r)$ is the density, $P=P(r)$ is the pressure, and $m=m(r)$ is the mass within theradial coordinate $r$. We solved these equations with the boundary conditions $\rho(0)=\rho_{\rm c}$, $m(0)=0$, $P(R_{\ast})=0$ and $m(R_{\ast})=M_{\ast}$. We employed AP4 \citep{AP4} as the EoS, $P=P(\rho)$. By trying other EoSs we have seen that our conclusions are independent of the choice of EoS.

In addition to $\rho(r)$, $P(r)$, and $m(r)$, we calculate the compactness
\begin{equation}
\eta (r) \equiv \frac{2Gm(r)}{rc^2},
\end{equation}
the Ricci scalar
\begin{equation}
{\cal R}(r)=\kappa(\rho c^2 -3 P), \qquad \kappa \equiv \frac{8\pi G}{c^4},
\label{ricci_scalar}
\end{equation}
the full contraction of the Ricci tensor
\begin{equation}
{\cal J}^2 \equiv {\cal R}_{\mu \nu} {\cal R}^{\mu \nu} = \kappa^2 \left[ (\rho c^2)^2 + 3P^2\right],
\label{ricci_tensor}
\end{equation}
the full contraction of the Riemann tensor (Kretschmann scalar)
\begin{align}
{\cal K}^2 &\equiv  {\cal R}^{\mu \nu \rho \sigma } {\cal R}_{\mu \nu \rho \sigma } \nonumber \\
&= \kappa^2 \left[ 3(\rho c^2)^2 + 3P^2 + 2P\rho c^2 \right] - \frac{16\kappa  G m \rho }{r^3 }
+ \frac{48 G^2 m^2}{r^6 c^4},
\label{K2}
\end{align}
and the full contraction of the Weyl tensor
\begin{equation}
{\cal W}^2 \equiv {\cal C}^{\mu \nu \rho \sigma }{\cal C}_{\mu \nu \rho \sigma } = \frac43 \left( \frac{6Gm}{c^2 r^3} -\kappa \rho c^2 \right)^2
\label{W2}
\end{equation}
within the star. The derivations of these curvature scalars for spherically symmetric metric in GR are presented in Appendix~\ref{app1}.

Which of these curvature scalars should one use as a measure of the curvature within a neutron star? It is well known that the components of the Ricci tensor ${\cal R}_{\mu \nu}$ and the Ricci scalar ${\cal R}$ vanish outside the star.  The vanishing of ${\cal R}_{\mu \nu}$ and ${\cal R}$  in vacuum can also be seen from Eqs. (\ref{ricci_scalar}) and (\ref{ricci_tensor}), as they depend only on $P$ and $\rho$ which vanish at the surface. This does not, however, mean that the spacetime of the vacuum as described by the Schwarzschild metric is not curved:
the Riemann tensor has nonvanishing components, e.g.\ $\tensor{{\cal R}}{^1_{010}}=-2GM_{\ast}/c^2 r^3$, in vacuum.
Thus the non-vanishing components of the Riemann tensor are more suitable measures of the curvature of the spacetime than the Ricci scalar and the components of the Ricci tensor. Instead of dealing with all such components---all of which are at the same order---one can employ the square root of the full contraction of the Riemann tensor (the Kretschmann scalar) ${\cal K} \equiv \sqrt{{\cal R}^{\mu\nu\alpha\beta}{\cal R}_{\mu\nu\alpha\beta}}=4\sqrt{3}GM_{\ast}/c^2 r^3$, as a measure of the curvature in vacuum. Note that in vacuum this is equal to the square root of the full contraction of the Weyl tensor, ${\cal W}$, as can easily be seen from Eqs. (\ref{K2}) and (\ref{W2}) with $m(r) \rightarrow M_{\ast}$ and $P=\rho=0$.  Within the star we employ ${\cal K}$ and ${\cal W}$ as two reasonable measures of the curvature. They attain different values within the star and approach each other at the crust, becoming identical in vacuum.

\section{Results}
\label{sec-results}

\begin{figure*}
\includegraphics[width=0.49\textwidth]{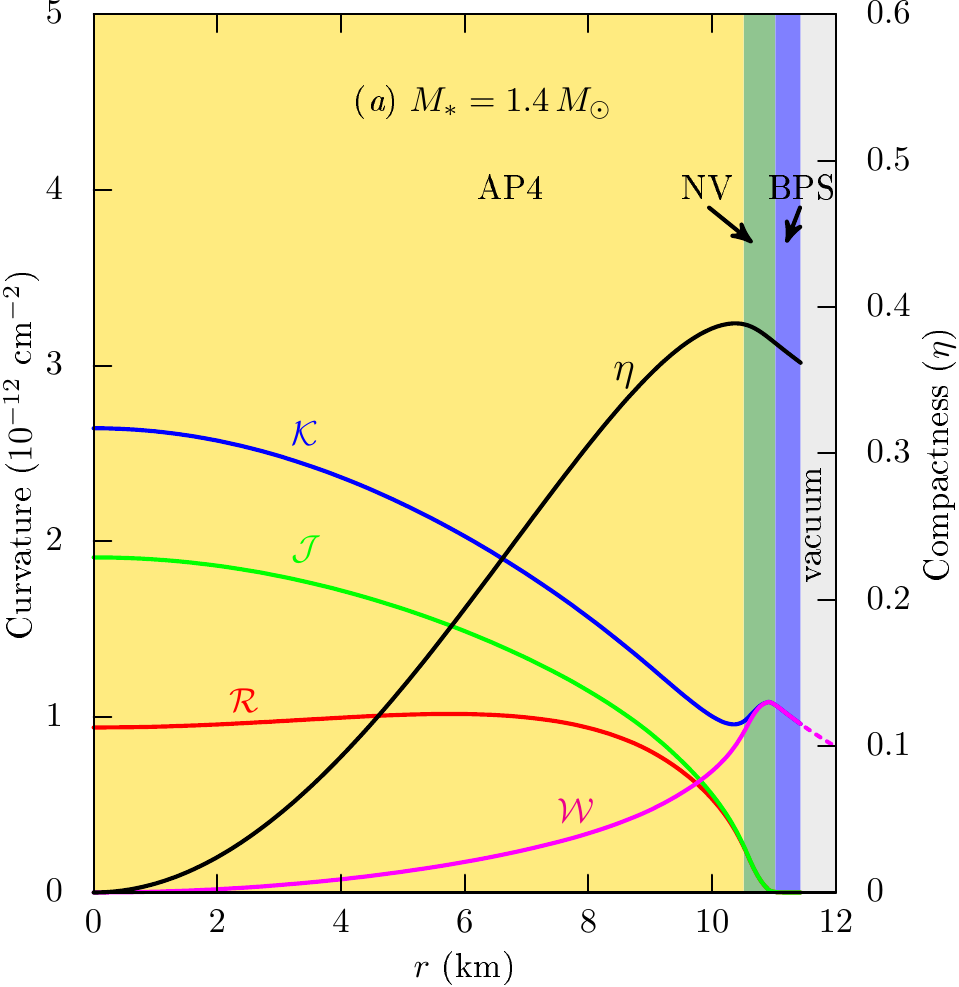}
\includegraphics[width=0.49\textwidth]{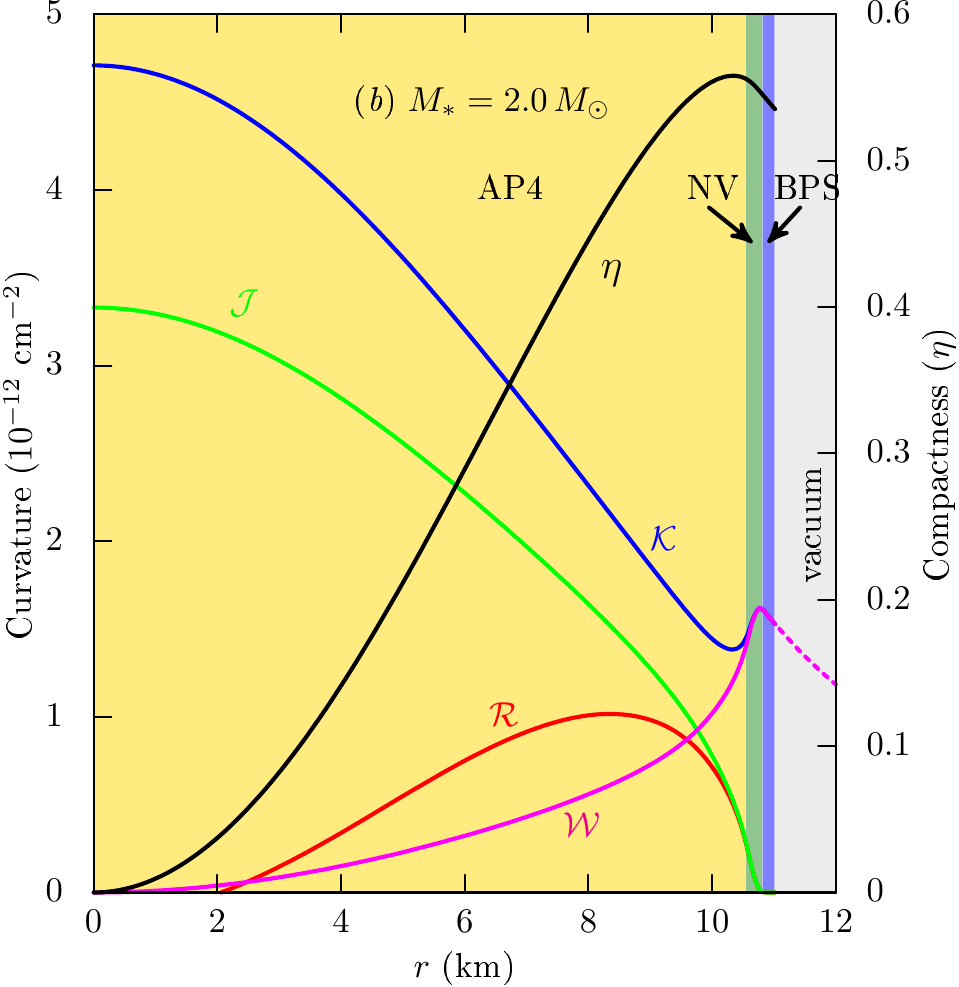}
\caption{The curvature scalars ${\cal R}$, ${\cal J}$, ${\cal K}$, and ${\cal W}$, and compactness $\eta$ within a neutron star. The left panel ({\it a}) is for a star with $M_{\ast}=1.4\,M_{\odot}$ and $R_{\ast}=11.43$ km, and the right panel ({\it b}) is for one with $M_{\ast}=2M_{\odot}$ and $R_{\ast}=11.01$ km in GR. The core (AP4), inner crust (NV) and outer crust (BPS) are shaded with different colors. The grey shaded region is vacuum at which ${\cal K}={\cal W}=4\sqrt{3}GM_{\ast}/c^2r^3$ is shown with dashed lines. At the centre ${\cal K}_{\rm c} \simeq 2.64\times 10^{-12}$~cm$^{-2}$ for $M_{\ast}=1.4\,M_{\odot}$ and ${\cal K}_{\rm c} \simeq 4.7\times 10^{-12}$~cm$^{-2}$ for $M_{\ast}=2\,M_{\odot}$. The curvatures ${\cal K} \simeq {\cal W} \sim 10^{-12}$ cm$^{-2}$ as well as the compactness $\eta \simeq 0.25$ at the crust are orders of magnitude larger than the curvature and compactness probed in the Solar System tests, ${\cal K}_{\odot} \sim 10^{-28}$ cm$^{-2}$ and $\eta_{\odot} \sim 10^{-5}$. It is seen in the right panel ({\it b}) that the Ricci scalar at the center of the star is negative up to $\simeq 2{\rm km}$. The fact that Ricci scalar takes negative values means that the AP4 EoS violates $\rho c^2 \geq 3P$ for high densities. This is an issue with some EoS that are not fully compatible with special relativity but that describe nuclear matter based on the nonrelativistic Schr\"{o}dinger equation rather than a relativistic mean-field theory.
}
\label{fig-curvature}
\end{figure*}

In Fig.~\ref{fig-curvature} we show the radial variation of the curvature scalars
${\cal R}$, ${\cal J}$, ${\cal K}$, and ${\cal W}$ together with the compactness $\eta$ within
a neutron star of ({\it a}) $M_{\ast}=1.4\,M_{\odot}$ and ({\it b}) $M_{\ast}=2\, M_{\odot}$.
We have used the AP4 EoS \cite{AP4} for the core, the NV EoS for the inner crust \citep{NV}, and the
BPS EoS for the outer crust \citep{BPS}. To obtain a $1.4\,M_{\odot}$ star we
employed a central density of $\rho_{\rm c} = 0.984\times 10^{15}$~g~cm$^{-3}$, which corresponds to a central pressure of
$P_{\rm c} = 1.44 \times 10^{35}$~dyne~cm$^{-2}$. To obtain a star with $2\,M_{\odot}$ we employed
$\rho_{\rm c} =1.53 \times 10^{15}$~g~cm$^{-3}$, which corresponds to
$P_{\rm c} = 4.85 \times 10^{35}$~dyne~cm$^{-2}$.
We see that at the center of the $1.4\,M_{\odot}$ and $2\,M_{\odot}$ stars, ${\cal K}_{\rm c} \simeq 2.64\times 10^{-12}$~cm$^{-2}$ and ${\cal K}_{\rm c} \simeq 4.7\times 10^{-12}$~cm$^{-2}$, respectively and
${\cal K}$ decreases only a few times towards the surface, remaining 15 orders of magnitude larger than ${\cal K}_{\odot}$.
If one uses ${\cal K}$ as a measure of curvature one would immediately be drawn
to the conclusion that GR is in a less tested regime than EoS  over the entire star.
In order to give the EoS uncertainty a ``chance'' to exceed that of gravity (at least at the inner parts of the star)
we choose the square root of the full contraction of the Weyl tensor ${\cal W}$---which vanishes at the center and increases towards the surface---as a measure of the curvature.

\begin{figure*}
\includegraphics[width=0.49\textwidth]{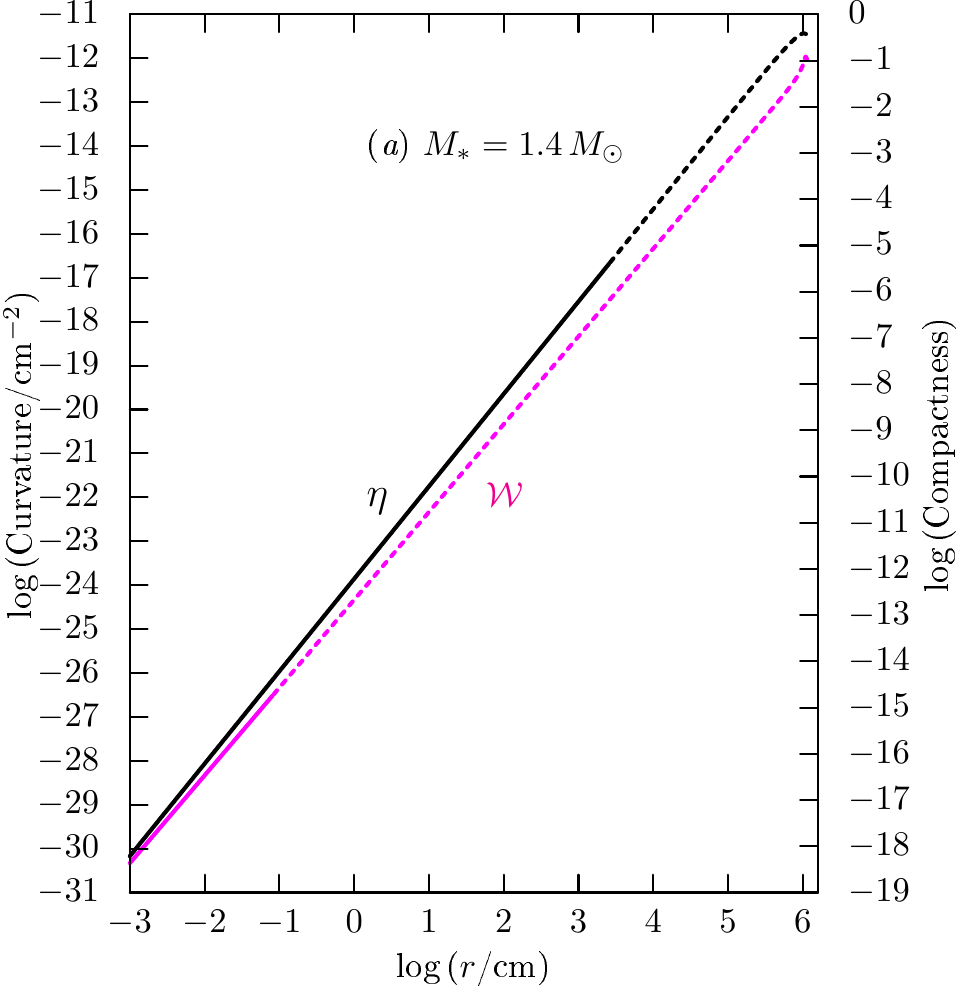}
\includegraphics[width=0.49\textwidth]{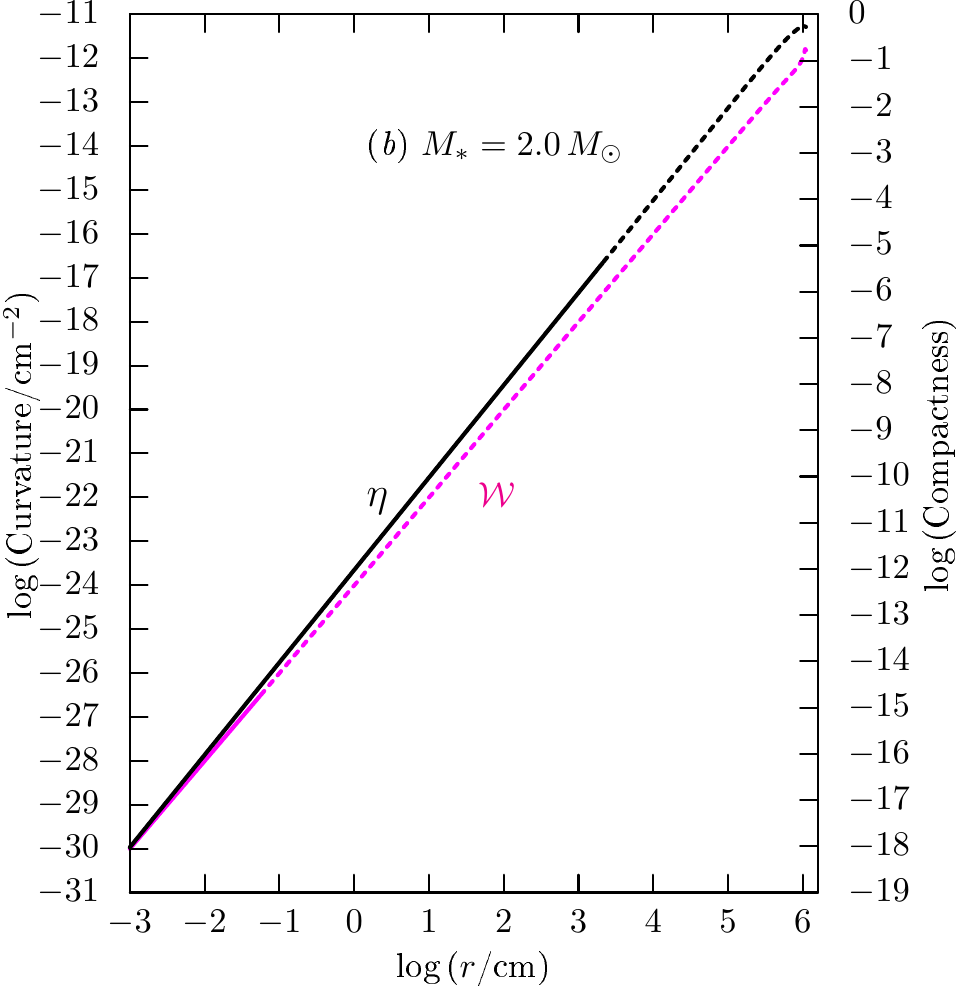}
\caption{The curvature scalar ${\cal W}$ and compactness $\eta$ within a neutron star.
Solid lines stand for the regime of gravitational strength probed in Solar System tests
(${\cal W}<{\cal W}_{\odot} = 3.06\times 10^{-27}$ cm$^{-2}$ and $\eta < \eta_{\odot} = 4.27 \times 10^{-6}$)
and dashed lines show the untested gravity strengths.
The left panel is for a star with $M_{\ast}=1.4\,M_{\odot}$ and $R_{\ast}=11.43$~km,
and right panel is for one with $M_{\ast}=2M_{\odot}$ and $R_{\ast}=11.01$~km in GR. We employed the AP4 EoS for the core,
NV for the inner crust and BPS for the outer crust.
The compactness is above the well-tested regime of gravity for $r \gtrsim 10$ m and the curvature
is orders of magnitude above the well-tested regime throughout the star, $r \gtrsim 1$ mm.
The curvature ${\cal W} \sim 10^{-12}$ cm$^{-2}$ and the compactness $\eta \simeq 0.5$ at
the crust are orders of magnitude larger than the curvature and compactness probed in
the Solar System tests.}
\label{fig-curv_log}
\end{figure*}

In Fig. ~\ref{fig-curv_log} we show the radial variation, on a logarithmic scale, of
the square root of the full contraction of the Weyl tensor ${\cal W}$ and compactness $\eta$  within
a neutron star of (\emph{a}) $M_{\ast}=1.4\,M_{\odot}$ and (\emph{b}) $M_{\ast}=2\, M_{\odot}$, obtained again with the AP4 EoS.
We see that the ``compactness'' $\eta$ is above the value probed in Solar System tests
($\eta_{\odot} \sim 10^{-5}$) for $r \gtrsim 10$~m and increases as a power law to
reach $\eta \simeq 0.5$ at the crust. Even more remarkable is the ``curvature'' ${\cal W}$ which goes beyond ${\cal W}_{\odot} = 3.06\times 10^{-27}$~cm$^{-2}$ for $r \simeq 1$~mm and increases as a power law (${\cal W} \propto r^2$) for a large part of the star (see Appendix \ref{app2}). As a result the curvature ${\cal W}$ is also several orders of magnitude larger throughout almost the entire star than the values in the Solar System at which GR is tested.


\section{DISCUSSION}
\label{sec-discuss}

Our results demonstrate that the strength of gravity within a neutron star is orders of magnitude larger than the gravitational strengths probed in Solar System tests, indicating that the presupposition of the validity of GR in determining the neutron star structure is an ``extravagant'' extrapolation. The EoS, on the other hand, is only an order of magnitude larger than what is being probed in nucleon scattering experiments. This does not only imply that an M-R measurement would constrain gravity models more than EoS
models, but also that---given that EoSs produce very different M-R relations---it would not be possible to investigate deviations from GR by any measurement of M-R unless the EoS is fully constrained by Earth-based experiments.
The strong gravity of neutron stars, nevertheless, might allow one to constrain modified gravity models in a perturbative approach (see, e.g., Refs. \citep[]{coo+10,ara+11,del+12}).

Our results are consistent with that of DeDeo and Psaltis \citep{ded03}, where the authors showed that M-R relations obtained by different models of gravity show a much greater variance than M-R relations obtained with different EoSs. Our results complement those of DeDeo and Psaltis \citep{ded03} by quantifying the strength of gravity \emph{within} the neutron stars and clarifies the fact that the domain where gravity is in a not-well-tested regime largely overlaps with the domain where the EoS is not constrained.

Our results for the domain of tested validity obviously depend on the gravity model chosen, i.e., GR.
The analysis we provide can be repeated for different viable gravity models with parameters
constrained  by Solar System tests. Though such a systematic survey is beyond the scope of
this paper, the present constraints on the mass and radius taken roughly as
$M_{\ast}=1-2\,M_{\odot}$ and $R=8-15$~km, already lead to a very large compactness and curvature at the surface.
These values obtained via phenomena outside the neutron star can be taken to be independent of the gravity model employed because almost all models of gravity yield the same vacuum solutions obtained with GR \citep{psa+08}.
It then seems obvious that in any gravity model the curvature and compactness within the neutron star should increase radially from zero to values that are orders of magnitude larger than those probed in the Solar System tests. Hence any model of gravity when employed for the structure of neutron stars would actually be extrapolated from its domain of tested validity.

A systematic stretching of GR beyond its regime of tested validity  by several orders of magnitude naturally raises the question of whether there is any bias in the historical development of EoSs, i.e., whether they have evolved towards producing reasonable M-R relations when used within the framework of GR. Such a question becomes more significant given that astrophysical measurements \citep{guv+1820,guv+1608,guv+1745,oze+10,guv13,oze+12,ste+10,gui+13}, except for that of Refs. \citep{sul+11} and \citep{neu+12}, yield radii systematically smaller than that implied by the Lead Radius Experiment (PREX) \citep{abr+12}  within the framework of GR \citep{fat12}.

GR sets a maximum mass $M_{\max}$ for neutron stars \citep{rho74} with a different value for each EoS (see Refs. \citep[]{cha+13,hae12} for a review). The ``two solar mass'' pulsars PSR J1614$-$2230 \citep{dem+10} and PSR J0348$+$0432 \citep{ant+13} suggest that $M_{\max} > 2 M_{\odot}$. The statistical analysis of K{\i}z{\i}ltan \emph{et al}.\ \citep{kiz+13} further suggests that the distribution of masses of neutron stars in nature are not cut off at the upper end such that stellar evolution could produce a low number of $\simeq 2.5\,M_{\odot}$ neutron stars. Within GR this favors stiff EoSs that to yield maximum allowable masses of neutron stars larger than 2.5 solar masses.
It is hard to reconcile this result with the small measured radii that favor soft EoSs within GR. This may signal deviations from GR in strong gravity unless there are systematic errors of order 30\% in the radius measurements.
We are yet unaware of any gravity model that yields 2.5 solar-mass objects with 9-11~km with the present EoSs that is also consistent with the Solar System tests. Recently, Arapo{\u g}lu \emph{et al}. \citep{ara+11} showed within an $f(R)=R+\alpha R^2$ gravity model that the maximum mass of neutron stars becomes larger for smaller values of $\alpha$, while the radius remains nearly constant. This model of gravity then gives large-mass objects for $\alpha<0$ even with soft EoSs. The nonrelativistic limit of such a gravity model, however, requires $\alpha>0$ \citep{naf10}, rendering solutions with $\alpha <0$ nonphysical.

Both the compactness and the curvature reach their maximum values near the crust.
Considering that the EoS of the crust is fairly
well known \citep{BPS,NV}, we are led to the conclusion that the crust of neutron stars provides the best sites in the
Universe for identifying deviations from GR in strong gravity. Such phenomena displayed by the crust are,
e.g., the glitch activity of young pulsars \citep{alp05} and modulation of the cooling of neutron stars \citep{yak04}.
One may then expect deviations
from GR to be identified from, e.g.,\ the glitch activity of pulsars given that GR is the least constrained physics input in the description of the phenomena by 14 orders of magnitude in curvature and 5 orders of magnitude in compactness, according to Fig.~\ref{fig-curv_log}. Such a deviation from GR may already have
been observed! The recent inference \citep{and+12} that the moment of inertia of the crust of neutron
stars is not sufficient to understand some of the glitch activity through the highly successful
vortex creep model \citep{alp+84,alp+93} may signal a  deviation from GR in strong gravity rather than a
requirement for the modification of the model itself. An investigation of the effect of scalar fields \citep{dam93}
on the moment of inertia of neutron stars is left for a future work.

\begin{center}
{\bf ACKNOWLEDGEMENTS}
\end{center}

\noindent We acknowledge support from the scientific
and technological council of Turkey (T\"{U}B\.{I}TAK)
with the project number 112T105. This work was supported by the COST Action  MP1304.

\onecolumngrid
\appendix

\section{Curvature scalars in the star}
\label{app1}
In this appendix we derive the curvature scalars inside a spherically symmetric mass distribution.
Specifically, we calculate the Ricci scalar ${\cal R}$, the full contraction of
the Ricci tensor ${\cal J}^2 \equiv {\cal R}^{\mu\nu}{\cal R}_{\mu\nu}$,
the full contraction of the Riemann tensor (namely the Kretschmann scalar)
${\cal K}^2 \equiv {\cal R}^{\mu\nu\alpha\beta}{\cal R}_{\mu\nu\alpha\beta}$ and the full contraction of the Weyl tensor ${\cal W}^2 \equiv {\cal C}^{\mu\nu\alpha\beta}{\cal C}_{\mu\nu\alpha\beta}$ in terms of pressure $P=P(r)$, density $\rho=\rho(r)$ and mass $m=m(r)$ within the radial coordinate $r$.

The tidal force is the only sign of gravity that cannot be cast aside by a coordinate transformation. The tidal force on a body moving along a geodesic leads to a distortion of the shape of the body and a change in the volume of the body, both of which are captured by the Riemann curvature tensor.
The trace component of the Riemann tensor, the Ricci curvature, conveys precisely the information about how volumes change in the presence of tidal forces. The Weyl tensor, being the traceless component of the Riemann tensor, only conveys information of how the shape of the body is distorted by the tidal force and nothing about how the volume of the body changes.

We choose the coordinates as ${t,r,\theta,\phi}$, respectively.
The most general spherically symmetric metric is given as follows:
\begin{equation}
{\rm d}s^{2}=-{\rm e}^{a}c^2{\rm d}t^{2}+{\rm e}^{b}{\rm d}r^{2}+r^{2}{\rm d}\theta^{2}+r^{2}\sin^2 \theta {\rm d}\phi^{2}
\label{metric}
\end{equation}
where $a=a(r)$ and $b=b(r)$. Here
\begin{equation}
g_{11}={\rm e}^{b}=\left(1-\frac{2Gm}{c^{2}r}\right)^{-1},
\label{g_11}
\end{equation}
but we refrain from using it until the end.

Einstein tensor with mixed components is
\begin{equation}
\tensor{G}{^{\mu}_{\nu}} \equiv \tensor{{\cal R}}{^{\mu}_{\nu}}-\frac{1}{2}\tensor{\delta}{^{\mu}_{\nu}}{\cal R}
\label{EinsteinTensor}
\end{equation}
where $\tensor{{\cal R}}{^\mu_\nu}$ is Ricci tensor with the mixed components.
The mixed components of the Einstein tensor are obtained from the metric as
\begin{align}
\tensor{G}{^{0}_{0}}&=-\frac{{\rm e}^{b}-1+rb^{\prime }}{r^{2}{\rm e}^{b}} \label{G00}  \\
\tensor{G}{^{1}_{1}}&=-\frac{{\rm e}^{b}-1-ra^{\prime }}{{\rm e}^{b}r^{2}} \label{G11}  \\
\tensor{G}{^{2}_{2}}&=-\frac{2(b^{\prime }-a^{\prime })-(2a^{\prime
\prime }-a^{\prime }b^{\prime }+\left( a^{\prime }\right) ^{2})r}{4r{\rm e}^{b}} \label{G22} \\
\tensor{G}{^{3}_{3}}&= \tensor{G}{^{2}_{2}} \label{G33}
\end{align}
where the primes denote derivatives with respect to $r$.

From the metric given in Eq. (\ref{metric}) we derive the Ricci scalar as
\begin{equation}
{\cal R}=\frac{\left[4r b^{\prime}-2a^{\prime \prime } + b^{\prime } a^{\prime
}-(a^{\prime })^2 \right]r^2   +4({\rm e}^{b}-1-r a^{\prime })}{2r^{2}{\rm e}^{b}},
\label{ricci}
\end{equation}
the full contraction of the Ricci tensor, ${\cal J}^2\equiv {\cal R}_{\mu \nu }{\cal R}^{\mu \nu }$ as
\begin{eqnarray}
{\cal J}^2=\frac{1}{8r^{4}{\rm e}^{2b}} &&\{ 16(1-{\rm e}^{b})^{2}+
16r(a^{\prime }-b^{\prime })(1-{\rm e}^{b})+4r^{2}\left[3(a^{\prime})^{2}+3(b^{\prime})^{2}-2b^{\prime }a^{\prime }\right] \nonumber\\
&&+4r^{3}\left[2a^{\prime\prime}a^{\prime}+(a^{\prime})^{3}-2a^{\prime\prime}b^{\prime}
+(b^{\prime})^{2}a^{\prime}-2(a^{\prime})^{2}b^{\prime}\right]  \nonumber\\
&&+ 4r^{4}\left[4a^{\prime\prime}(a^{\prime})^{2}+(a^{\prime})^{2}(b^{\prime})^{2}-2b^{\prime}(a^{\prime})^{3}+
4(a^{\prime\prime})^{2}
+(a^{\prime})^{4}-4a^{\prime\prime}a^{\prime}b^{\prime}\right] \},
\label{riccisq}
\end{eqnarray}
the Kretschmann scalar, ${\cal K}^2 \equiv {\cal R}_{\mu \nu \rho \sigma }{\cal R}^{\mu \nu \rho \sigma }$ where ${\cal R}_{\mu \nu \rho \sigma }$ is Riemann curvature tensor, as
\begin{eqnarray}
{\cal K}^2=\frac{1}{4r^{4}{\rm e}^{2b}}&&\{16(1-{\rm e}^{b})^{2}+8r^{2}\left[(a^{\prime})^{2}+(b^{\prime})^{2}\right]
\nonumber\\
&&+r^{4} \left[4(a^{\prime\prime})^{2}+4a^{\prime\prime}(a^{\prime})^{2}-4a^{\prime\prime}a^{\prime}b^{\prime}+
(a^{\prime})^{4}-2b^{\prime}(a^{\prime})^{3}+(a^{\prime})^{2}(b^{\prime})^{2}\right] \}
\label{riemannsq}
\end{eqnarray}
\cite{bro03}, and the full contraction of the Weyl tensor, ${\cal W}^2 \equiv {\cal C}^{\mu \nu \rho \sigma }{\cal C}_{\mu \nu \rho \sigma }$, as
\begin{eqnarray}
{\cal W}^2 &=&\frac{1}{3}\left(\frac{r^{2}[-2a^{\prime\prime}-(a^{\prime})^{2}+a^{\prime}b^{\prime}]
+2r(a^{\prime}+b^{\prime})+4({\rm e}^{b}-1)}{2r^{2}{\rm e}^{b}}\right)^{2}.
\label{weylsq}
\end{eqnarray}

We solve $a'$, $a''$ and $b'$ from Eqs. (\ref{G00}), (\ref{G11}) and (\ref{G22}) to obtain
\begin{eqnarray}
a^{\prime } &=&\frac{ \tensor{G}{^{1}_{1}}r^{2}{\rm e}^{b}+{\rm e}^{b}-1}{r}   \label{aprime}  \\
b^{\prime } &=&-\frac{ \tensor{G}{^{0}_{0}}r^2{\rm e}^{b}+e^{b}-1}{r}   \label{bprime}  \\
a^{\prime \prime } &=&\frac{2( \tensor{G}{^{2}_{2}} + \tensor{G}{^{3}_{3}})r{\rm e}^{b}
+(2 +r a^{\prime })( b^{\prime }-a^{\prime }) }{2r}  \nonumber \\
&=& -\frac{r^2{\rm e}^{b}\left( \tensor{G}{^0_0} - \tensor{G}{^1_1}-2\tensor{G}{^2_2}-2 \tensor{G}{^3_3}\right)
+r^2 {\rm e}^{2b}\left(3 \tensor{G}{^1_1}+\tensor{G}{^0_0}\right)+2\left({\rm e}^{2b}-1\right)
+r^4{\rm e}^{2b}\tensor{G}{^1_1}\left( \tensor{G}{^0_0} +\tensor{G}{^1_1}\right)
}{2r^{2}}
\label{adoubleprime}
\end{eqnarray}
where in the last step we used $b'-a'$ obtained from the first two equations.
Plugging these into Eq. (\ref{ricci}) we obtain the Ricci scalar in terms of the components
of the Einstein tensor as
\begin{eqnarray}
{\cal R} &=& -\tensor{G}{^{0}_{0}}- \tensor{G}{^{1}_{1}} -\tensor{G}{^{2}_{2}}- \tensor{G}{^{3}_{3}} \nonumber \\
  &=& -\tensor{G}{^{\mu}_{\mu}}.  \label{ricci1}
\end{eqnarray}
Similarly, by plugging Eq.s (\ref{aprime})-(\ref{adoubleprime}) into Eq. (\ref{riccisq}) we obtain
\begin{eqnarray}
{\cal R}_{\mu \nu }{\cal R}^{\mu \nu }
&=&\left( \tensor{G}{^{0}_{0}}\right)^2 + \left( \tensor{G}{^{1}_{1}}\right)^2+\left(\tensor{G}{^{2}_{2}}\right)^2 +\left( \tensor{G}{^{3}_{3}}\right)^2 . \label{riccisquare}
\end{eqnarray}
Again, by plugging Eqs. (\ref{aprime})-(\ref{adoubleprime}) into Eq. (\ref{riemannsq}) the Kretschmann scalar can be expressed as
\begin{eqnarray}
{\cal K}^2 &=&
2\tensor{G}{^{0}_{0}}\tensor{G}{^{1}_{1}}-2\tensor{G}{^{0}_{0}}\tensor{G}{^{2}_{2}}-2\tensor{G}{^{0}_{0}}\tensor{G}{^{3}_{3}}-
2\tensor{G}{^{2}_{2}}\tensor{G}{^{1}_{1}}-2\tensor{G}{^{2}_{2}}\tensor{G}{^{3}_{3}}\nonumber \\
&&+3\left( \tensor{G}{^{0}_{0}}\right)^{2}+
3\left( \tensor{G}{^{1}_{1}}\right) ^{2}+2\left( \tensor{G}{^{2}_{2}}\right) ^{2}+2\left(
\tensor{G}{^{3}_{3}}\right) ^{2}   \nonumber\\
&&+8\left( \tensor{G}{^{0}_{0}}+\tensor{G}{^{1}_{1}}-\frac{1}{2}(\tensor{G}{^{2}_{2}}+\tensor{G}{^{3}_{3}})\right)\left( 1-{\rm e}^{-b}\right)
r^{2}+12\left(1-{\rm e}^{b}\right)^{2}{\rm e}^{-2b}. \label{RiemannSquare}
\end{eqnarray}
Finally, by plugging Eqs. (\ref{aprime})-(\ref{adoubleprime}) into Eq. (\ref{weylsq})
the full contraction of the Weyl tensor becomes
\begin{equation}
{\cal W}^2 = \frac43 \left[   \tensor{G}{^{0}_{0}}+\tensor{G}{^{1}_{1}}-\frac{1}{2}(\tensor{G}{^{2}_{2}}+\tensor{G}{^{3}_{3}}) + \frac{3({\rm e}^{b}-1)}{r^{2}{\rm e}^{b}}\right]^2 . \label{WeylSquare}
\end{equation}

The Einstein field equation with mixed components is
\begin{equation}
\tensor{G}{^{\mu}_{\nu}}=\kappa \tensor{T}{^{\mu}_{\nu}}
\label{EinsteinField}
\end{equation}
where $\tensor{T}{^{\mu}_{\nu}}$ are the mixed components of the energy-momentum tensor
and  $\kappa \equiv 8\pi G/c^4$.
We assume that the energy-momentum tensor is that of a perfect fluid,
\begin{equation}
 \tensor{T}{^{\mu\nu}}=\left(\rho c^2 + P \right) \tensor{u}{^{\mu}} \tensor{u}{^{\nu}} + P \tensor{g}{^{\mu\nu}}
\end{equation}
where $\rho$ is the density, $P$ is the pressure and $u^{\mu}$ is four-velocity.
Contracting the $\tensor{T}{^{\mu\nu}}$ with $ \tensor{g}{_{\mu\nu}}$, the mixed components of the energy-momentum tensor can be written as
\begin{eqnarray}
\tensor{T}{^{\mu}_{\nu}}=\left(\rho c^2 + P \right)u^{\mu}u_{\nu} + \tensor{\delta}{^\mu_\nu} P.
\label{MixedEmt}
\end{eqnarray}
Thus $\tensor{T}{^0_0}=-\rho c^2$ and $\tensor{T}{^i_i}=P$. Using $u^\mu u_\mu = -1$ and $\tensor{\delta}{^\mu_\mu}=4$ the trace of the energy-momentum equation becomes
\begin{equation}
\tensor{T}{^{\mu}_{\mu}}=-\left(\rho c^2 -3 P \right).
\label{trace}
\end{equation}

Equation (\ref{ricci1}) and the Einstein equation (\ref{EinsteinField}) imply ${\cal R}=-\kappa \tensor{T}{^{\mu}_{\mu}}$, and by referring further to Eq. (\ref{trace}) the Ricci scalar is obtained as
\begin{equation}
{\cal R} = \kappa \left(\rho c^2-3 P\right)
\end{equation}
in terms of the density $\rho$, and pressure $P$ within the star. Similarly, the full contraction of the Ricci tensor in Eq. (\ref{riccisquare})
becomes
\begin{equation}
{\cal R}_{\mu \nu }{\cal R}^{\mu \nu } =  \kappa^2 \left[ (\rho c^2)^2 + 3P^2\right].
\end{equation}

Using the components of the mixed energy-momentum tensor given in Eq. (\ref{MixedEmt}) and $g_{11}$ given in Equation (\ref{g_11}) we obtain the  Kretschmann scalar given in Eq. (\ref{RiemannSquare}) as
\begin{equation}
{\cal K}^2 =\kappa^2 \left[ 3(\rho c^2)^2 + 3P^2 + 2P\rho c^2 \right] - \kappa \frac{16Gm}{r^3c^2}\rho c^2 + \frac{48G^2m^2}{r^6c^4},
\label{KK}
\end{equation}
and the full contraction of the Weyl tensor given in Eq. (\ref{WeylSquare}) as
\begin{equation}
{\cal W}^2 = \frac43 \left(\frac{6Gm}{c^2 r^3} - \kappa \rho c^2\right)^2 .
\label{WW}
\end{equation}

Note that at the surface of the star $\rho=0$, $P=0$ and $m=M_{\ast}$ and so one obtains ${\cal K}^2={\cal W}^2=48G^2M_{\ast}^2/c^4 r^6$ outside the star, a well-known result for the Schwarzschild metric. As a further check, note that
these satisfy:
\begin{equation}
{\cal K}^2 = {\cal W}^2 + 2{\cal J}^2-\frac13 {\cal R}^2
\end{equation}
(see Eq. (3) in Ref. \citep[]{che02}).

For a star with uniform density $m=\frac43 \pi r^3 \rho$ and plugging this into Eq.(\ref{WW})
 the full contraction of the Weyl tensor vanishes everywhere within the star.
 This means that ${\cal W}^2=0$ at the center of the star because $m\rightarrow \frac43 \pi r^3 \rho_{\rm c}$ as $r\rightarrow 0$. It also means that ${\cal W}^2>0$ everywhere within a realistic star stratified as ${\rm d}P/{\rm d}r<0$.

For a uniform-density star the negative term in Eq. (\ref{KK}) becomes
$-\frac83 \kappa^2 (\rho c^2)^2$ which is less than the first term $3 \kappa^2 (\rho c^2)^2$, meaning that ${\cal K}^2$ is also positive definite. We have seen that the Ricci scalar ${\cal R}$ changes sign within the star only for massive neutron stars $M \gtrsim 2\, M_{\odot}$ for the AP4 EoS. We observe that ${\cal K}$ and ${\cal J}$ attain their maximum values at the center of the star. Near the surface ${\cal J}$ and ${\cal R}$ vanish, as expected as their vacuum value is zero. Within the crust ${\cal W}$ and ${\cal K}$ approach each other, and they both have a local maximum and drop slightly to match $4\sqrt{3}GM_{\ast}/R_{\ast}^3 c^2$ at the surface.

\section{Calculating $\cal{W}$ near the origin}
\label{app2}

The full contraction of the Weyl tensor ${\cal W}^2 \equiv {\cal C}^{\mu \nu \rho \sigma }{\cal C}_{\mu \nu \rho \sigma }$ given in Eqn. (\ref{WW}) vanishes at the origin. We found that the numerical calculation  of
\begin{equation}
{\cal W} = \frac{2}{\sqrt{3}}  \left(\frac{6Gm}{c^2 r^3} - \kappa \rho c^2\right)
\label{W}
\end{equation}
near the origin ($r<1$~cm) requires careful treatment because the two terms in the parentheses are so close to each other that their subtraction results in truncation errors larger than the value itself. We have solved the issue by employing a series expansion of ${\cal W}$ near the center.

Consider the series expansion of $P$, $\rho$ and $m$ near the origin,
\begin{align}
P &= P(0)+P'(0)r + P''(0)\frac{r^2}{2} + O(r^3) \\
\rho &= \rho(0)+\rho'(0)r +  \rho''(0) \frac{r^2}{2} + O(r^3) \\
m &= m(0)+m'(0)r + m''(0)\frac{r^2}{2} +  m'''(0)\frac{r^3}{6}
+ m^{(iv)}(0) \frac{r^4}{24} + m^{(v)}(0)\frac{r^5}{120} +O(r^6)
\end{align}
Note that $P(0)=P_{\rm c}$, $\rho(0)=\rho_{\rm c}$, and $m(0)=0$ are used as boundary conditions. According to the TOV equation given in Eq. (\ref{tov2}), i.e.,\ $m'=4\pi r^2 \rho$, we have further that $m'(0)=0$.
Taking the higher derivatives,
\begin{align}
m'' &= 4\pi (2r\rho + r^2 \rho'),  & m''(0) &= 0  \\
m''' &= 4\pi (2\rho + 4r\rho' + r^2 \rho''), &  m'''(0) &= 8\pi \rho_{\rm c} \\
m^{(iv)} &= 4\pi (6\rho'  + 6r\rho''  +  r^2 \rho'''),  & m^{(iv)}(0) &= 24\pi \rho'(0)  \\
m^{(v)} &= 4\pi (12\rho''   + 8r\rho'''  +  r^2 \rho^{(iv)}), &  m^{(v)}(0) &= 48\pi \rho''(0)
\end{align}
Thus to a first order $m=\frac43 \pi r^3 \rho_{\rm c}$ near the origin, and by plugging this into Eq. (\ref{tov1}) $P'(0)=0$.

The first TOV equation can be written as $P'=Gm\rho \xi/r^2$, where $\xi$ is the relativistic correction
\begin{equation}
\xi =  \left( 1+ \frac{P}{\rho c^2}\right)
                 \left(1+\frac{4\pi r^3 P}{mc^2} \right)
                 \left(1-\frac{2Gm}{rc^2}\right)^{-1}
\end{equation}
The first TOV equation, by the chain rule
\begin{equation}
\frac{{\rm d}P}{{\rm d}r}=\frac{{\rm d}P}{{\rm d}\rho}\frac{{\rm d}\rho}{{\rm d}r},
\end{equation}
becomes
\begin{equation}
\rho' = - \frac{Gm \rho^2 \xi}{\gamma r^2 P}
\label{rho1}
\end{equation}
where $\gamma \equiv {\rm d}\ln P / {\rm d} \ln \rho$. This implies $\rho'(0)=0$ upon using $m=\frac43 \pi r^3 \rho_{\rm c}$.

Taking the derivative of the TOV equation one obtains
\begin{equation}
P''=-\frac{Gm'\rho \xi}{r^2}-\frac{Gm\rho' \xi}{r^2}-\frac{Gm\rho \xi'}{r^2}+\frac{2Gm\rho \xi}{r^3}.
\end{equation}
Using the second TOV equation, $m'=4\pi r^2 \rho$, in the first term, noting that $\rho'(0)=0$ in the second term and using $m=\frac43 \pi r^3 \rho_{\rm c}$ for the third and forth terms one obtains
$P''(0) = -\frac{4\pi}{3}G \rho_{\rm c}^2 \xi_{\rm c}$ where $\xi_{\rm c}=(1+P_{\rm c}/\rho_{\rm c} c^2)(1+3P_{\rm c}/\rho_{\rm c} c^2)$.
Similarly, by taking the derivative of Eq.~(\ref{rho1}) one obtains
$\rho''(0) = -(4\pi/3)G\rho_{\rm c}^3 \xi_{\rm c} / \gamma_{\rm c} P_{\rm c}$.

Thus, near  the origin we have
\begin{align}
P &= P_{\rm c} -\frac{2\pi}{3}G \rho_{\rm c}^2 r^2 \xi  \\
\rho &= \rho_{\rm c}  +  \rho''(0) \frac{r^2}{2} \\
m &= \frac43 \pi r^3 \rho_{\rm c}  + \frac{2}{5} \pi \rho''(0) r^5
\end{align}
by which we can obtain
\begin{equation}
{\cal W}(r\rightarrow 0) = -\frac{2}{5\sqrt{3}} \kappa \rho''(0) c^2 r^2
\end{equation}
Here $\gamma_{\rm c} \simeq 3$ (for AP4) depending on the central density, which can be calculated from the EoS by $\gamma_{\rm c} = (\rho_{\rm c}/P_{\rm c})\Delta P/\Delta \rho$.

\end{document}